\documentclass[RNAAS]{aastex63}

\graphicspath{{./}}
\begin{document}

\title{TESS Science Processing Operations Center FFI Target List Products}

%% Note that the corresponding author command and emails has to come
%% before everything else. Also place all the emails in the \email
%% command instead of using multiple \email calls.
\correspondingauthor{Douglas A. Caldwell}
\email{dcaldwell@seti.org}

\author[0000-0003-1963-9616]{Douglas A. Caldwell}
\affiliation{SETI Institute}
\affiliation{NASA Ames Research Center}

\author[0000-0002-1949-4720]{Peter Tenenbaum}
\affiliation{SETI Institute}
\affiliation{NASA Ames Research Center}

\author[0000-0002-6778-7552]{Joseph D. Twicken}
\affiliation{SETI Institute}
\affiliation{NASA Ames Research Center}

\author[0000-0002-4715-9460]{Jon M. Jenkins}
\affiliation{NASA Ames Research Center}

\author[0000-0002-8219-9505]{Eric Ting}
\affiliation{NASA Ames Research Center}

\author[0000-0002-6148-7903]{Jeffrey C. Smith}
\affiliation{SETI Institute}
\affiliation{NASA Ames Research Center}

\author[0000-0002-3385-8391]{Christina Hedges}
\affiliation{Bay Area Environmental Research Institute}
\affiliation{NASA Ames Research Center}

\author[0000-0002-9113-7162]{Michael M. Fausnaugh}
\affiliation{Kavli Institute for Astrophysics and Space Science, Massachusetts Institute of Technology}

\author[0000-0003-4724-745X]{Mark Rose}
\affiliation{NASA Ames Research Center}

\author{Christopher Burke}
\affiliation{Kavli Institute for Astrophysics and Space Science, Massachusetts Institute of Technology}

%% Note that RNAAS manuscripts DO NOT have abstracts.
% They do now!  150 word abstract
\begin{abstract}
    We report the delivery to the Mikulski Archive for Space Telescopes of target pixel and light curve files for up to 160,000 targets selected from full-frame images (FFI) for each TESS Northern hemisphere observing sector. The data include calibrated target pixels, simple aperture photometry flux time series, and presearch data conditioning corrected flux time series. These data provide TESS users with high quality, uniform pipeline products for a selection of FFI targets, that would otherwise not be readily available. Additionally, we deliver cotrending basis vectors derived from the FFI targets to allow users to perform their own systematic error corrections. The selected targets include all 2-minute targets and additional targets selected from the TESS Input Catalog with a maximum of 10,000 targets per sector on each of the sixteen TESS CCDs. The data products are in the same format as the project-delivered files for the TESS 2-minute targets. 
\end{abstract}

%% Start the main body of the article. If no sections in the 
%% research note leave the \section call blank to make the title.

% count words with texcount
% texcount -v3 -merge -incbib -dir -sub=none -utf8 -sum rnaas.tex

\section{} 

The TESS mission \citep{rickerTESS} has observed most of the sky over the past two years with two data collection modes: 2-minute sampling of 20,000 targets for most sectors, and 30-minute sampled full-frame images (FFI) covering a 24$^\circ$ x 96$^\circ$ region of the sky in each sector. 
Since the start of the TESS Mission, the TESS Science Processing Operations Center (SPOC) pipeline has been used to calibrate FFIs and to assign world-coordinate system information to the FFI data delivered to the Mikulski Archive for Space Telescopes (MAST). 
The SPOC pipeline has generated target pixel files, light curves, and associated products from two--minute target data, but not from FFIs \citep{SPOC}. 
In the second year of the mission, the SPOC began processing additional 30--minute targets selected from the FFIs to create target pixel and light curve files for up to 160,000 targets per sector. 
By defining a set of target stars selected from the calibrated FFIs, we can flow the data for these targets through the SPOC pipeline to produce the same data products as for the two--minute targets: calibrated target pixel files, simple aperture photometry (SAP) flux time series \citep{PA2010, PA}, and presearch data conditioning (PDC) corrected flux time series \citep{smithSsMAP, stumpeMsMAP}. 
In addition, the pipeline generates a set of 30-minute sampled cotrending-basis vectors suitable for cotrending other target light curves generated from the FFIs.

We report here on the delivery to the MAST of these TESS-SPOC High-Level Science Products (HLSP) for FFI targets from Sectors~14 -- 26. HLSP data from the Southern hemisphere, Sectors~1 -- 13, will be delivered as they are reprocessed, as will data from the Extended Mission Sectors~27 and beyond. All of the TESS-SPOC FFI light curves, target pixel files, and cotrending basis vectors are available at the MAST as a High Level Science Product via
\dataset[10.17909/t9-wpz1-8s54]{\doi{10.17909/t9-wpz1-8s54}}.

\begin{figure}[ht]
\begin{center}
\includegraphics[scale=0.75,angle=0]{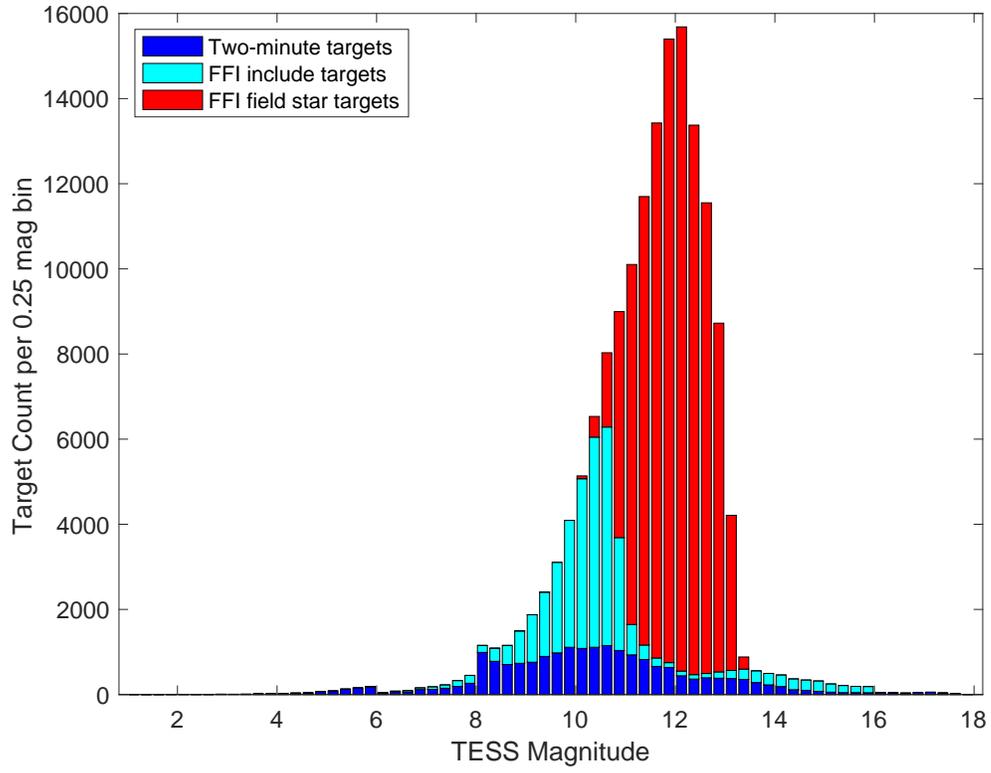}
\caption{Distribution of selected FFI targets from TESS observing Sector-14 shown as a stacked histogram. The FFI target list includes 156,217 total targets with 20,000 targets from the 2-minute list (blue), 31,175 FFI include targets (cyan), and 105,042 FFI field star targets selected by magnitude (red). \label{fig:1}}
\end{center}
\end{figure}

The philosophy behind our FFI target selection was to use the simplest approach possible within the constraints of the pipeline infrastructure design, while still selecting a set of targets that could prove useful for diverse scientific goals. In order to minimize the impact to operations, the FFI target lists need to be generated automatically based on criteria available to the pipeline, including the TESS Input Catalog \citep[TIC --][]{Stassun2019} and the pipeline-calculated crowding metric \citep{TAD}. FFI target selection is done in three steps:
\begin{enumerate}
    \item Select all \emph{two--minute} targets (nominally 20,000 per sector)
    \item Select potentially high value \emph{FFI include} targets with 
    \begin{itemize}
        \item H magnitude $\leq 10$ or distance $\leq 100$~pc (to include IR bright or nearby stars, which are valuable for exoplanet follow-up)
        % \item log surface gravity $\ge 1$
        \item crowding metric $\ge 0.5$ (at least 50\% of the flux in the photometric aperture is from the selected target)
        \item TESS magnitude $\leq 16$ (only TIC objects brighter than 16 are nominally available to the pipeline modules)
    \end{itemize}
    \item Select additional \emph{FFI field star} targets in order of TESS magnitude with 
    \begin{itemize}
        \item TESS magnitude $\leq 13.5$
        \item log surface gravity $\ge 3.5$ (select dwarfs and sub-giants)
        \item crowding metric $\ge 0.8$ (select stars that dominate the flux in their aperture)
    \end{itemize}
\end{enumerate}
Early testing indicated that there would be minimal impact to the processing time for a sector as long as the number of targets was limited to $\sim 10,000$ per CCD. To meet this limit, targets are selected on each CCD in the order given above until either 10,000 targets are selected, or all targets on that CCD meeting the third criterion have been selected. The number of FFI targets for the TESS Northern hemisphere sectors (14--26) has ranged from a low of 152,359 (Sector-22) to the limit of 160,000 (Sectors 17, 18, 19, 20, 21, 24, 25 26). The fixed target number limit per CCD results in a varying magnitude cutoff over the focal plane because of crowding variations across the large TESS field of view. The distribution of selected targets by magnitude for Sector~14 is given in Figure~\ref{fig:1}.

%% Figure: Sector 23 target magnitude histogram

%Articles can be submitted in \latex\ (preferably with the new "RNAAS"
%style option in AASTeX v6.2), MS/Word, or via the direct submission in the
%\href{http://www.authorea.com}{Authorea} or
%\href{http://www.overleaf.com}{Overleaf} online collaborative editors.

\acknowledgments

This paper is based on data collected by the TESS mission. Funding for the TESS mission is provided by the NASA Explorer Program. Resources were provided by the NASA High-End Computing (HEC) Program through the NASA Advanced Supercomputing (NAS) Division at Ames Research Center for the production of SPOC data products and these HLSP.

\facilities{\textit{TESS}}

\bibliography{tessFtl}{}
\bibliographystyle{aasjournal}

\end{document}